\documentstyle[preprint,aps]{revtex}

\begin{document}
\draft

\title{Effects of a torsion field on Big Bang nucleosynthesis}

\author{M. Br\"uggen\footnote{email: marcus@mpa-garching.mpg.de, phone: +49 89 3299 3245}
} 

\address{Max-Planck-Institut f\"ur Astrophysik, Karl-Schwarzschild-Str.1, 85740 Garching, Germany; and Churchill College, Storey's Way, Cambridge CB3 0DS, United Kingdom}

\date{\today}
\maketitle

\begin{abstract}
In this paper it is investigated whether torsion, which arises
naturally in most theories of quantum gravity, has observable
implications for the Big Bang nucleosynthesis. Torsion can lead to
spin flips amongst neutrinos thus turning them into sterile
neutrinos. In the early Universe they can alter the helium abundance
which is tightly constrained by observations. Here I calculate to what
extent torsion of the string theory type leads to a disagreement with
the Big Bang nucleosynthesis predictions.
\end{abstract}
\vspace{0.5cm}

\noindent keywords: Big Bang nucleosynthesis -- theories of gravitation --
nonstandard neutrino physics \\
\pacs{04.50.+h,14.60.St,98.80.Cq,98.80.Es}

%\pagebreak 
\section{Introduction}

In this paper I investigate the spin flip of neutrinos due to a
torsion field. Torsion fields arise naturally in most quantum theories
of gravitation. In these theories the spin of a particle is related to
the torsion in the same manner as the mass is related to the
curvature. Motivated by the work of Yang and Mills \cite{YaMi}, Kibble
\cite{Ki} and Sciama \cite{Sc} developed a gauge theory of gravity
that contains torsion as a necessary ingredient. More recently,
torsion fields have appeared in classical string theory \cite{KaRa},
supergravity theories \cite{Cr,Pa} and twistor theories of gravity
\cite{Ho}. Torsion is defined as the antisymmetric part of the affine
connection and introduces an additional coupling term in the Dirac
equation which can cause spin flips. Hammond \cite{Ha1} derived a
theory of gravity based on the assumption that torsion can be derived
from an antisymmetric potential. It was found that this leads to a
coupling which is of the same type as that obtained from string
theory.\\

To avoid confusion I should stress that spin flips can also be caused by
the interaction between spin and curvature. This is a separate
mechanism related to the frame-dragging effect in standard General
Relativity. The flipping of the neutrino's helicity in the vicinity of
rotating proto-neutron stars has been treated, e.g., in Ref. \cite{Br}.\\

Capozziello {\it et al.} \cite{Ca} have described the spin flip in a
semiclassical formalism by introducing a torsion term into the
Hamiltonian. They speculate that spin flips due to torsion could
become significant in the early Universe with profound
implications. In this paper I calculate the spin flip probability
due to torsion in the early Universe and investigate its consequences
on Big Bang nucleosynthesis (BBN). The effect of torsion on the
structure formation in cosmology has been calculated in
Ref. \cite{CaSt}.\\

\section{Spin flip due to torsion}

The spin-flip cross section in a torsion field of the string theory
type has been derived in Ref. \cite{Ha2}, and I will only briefly
sketch the derivation here. Details can be found {\it ibid}.
Torsion is defined as the antisymmetric part of the affine connection,
i.e. $\Gamma_{[\mu\nu]}^{\sigma}=S_{\mu\nu}^{\sigma}$, where
$\Gamma_{\mu\nu}^{\sigma}$ is the Christoffel symbol.\\

To derive the corresponding coupling in the Dirac equation one can
rewrite the connection in terms of the tetrads $e_{\mu}^a$, where the
Latin indices refer to the locally inertial frame and Greek indices to
a generic non-inertial frame. The nonholonomic index $a$ labels the
tetrad, while the holonomic index $\mu$ label the components of a
given tetrad. In nonholonomic coordinates the connection can be
expressed as

\begin{equation}
\Gamma_{abc}=-\Omega_{abc}+\Omega_{bca}-\Omega_{cab}+S_{abc} \label{Omega},
\end{equation}
where $\Omega^c_{\alpha\beta}=e_{[\alpha,\beta]}^c$ and
$\Omega^c_{ab}=e_{a}^{\alpha}e_b^{\beta}e^c_{\sigma}\Omega^{\sigma}_{\alpha\beta}$.\\

Assuming that torsion dominates gravitation, one can neglect the
$\Omega_{abc}$ (curvature) terms in Eq.\, (\ref{Omega}), and the Dirac
equation can be written in the form

\begin{equation}
\gamma^{\mu}\psi_{,\mu}
+\frac{imc}{\hbar}\psi={\textstyle\frac{1}{4}}S_{\mu\nu\sigma}\gamma^{\mu}\gamma^{\nu}\gamma^{\sigma}\psi,
\label{Dir}
\end{equation}
where $c$ denotes the speed of light and $m$ mass. Subsequently, the
scattering cross section in the high energy limit ($E\gg m$) can be
calculated by solving Eq.\,(\ref{Dir}) in the Born approximation.
Assuming a minimally coupled Lagrangian, the cross section for a
neutrino with mass $m$ and spin $S$ to undergo a spin flip by
scattering from a fixed particle is given by \cite{Ha2}:

\begin{equation}
\sigma \simeq 8.28\pi\left ( \frac{9 G S m}{4\hbar c^2}\right )^2 \label{sigma},
\end{equation}
where $G$ is the gravitational constant.

\section{Big Bang nucleosynthesis}

The striking agreement between the predictions of primordial
nucleosynthesis with the observed light element abundances is regarded
as a big success of Big Bang nucleosynthesis \cite{KoTu}. Thus the
predictions of BBN have become a stringent test for theories of
cosmology and particle physics.\\

The observed $^4$He abundance of $Y=0.235\pm 0.01$ \cite{KoTu}
constrains the number of neutrino species (active and sterile) to
$N_{\nu}<3.6$. When additional neutrino species are introduced in the
early Universe, the Universe would expand faster due to the increased
energy density, which in turn would lead to a higher neutron to proton
ration and hence to a higher helium yield. The effect of neutrino
mixing between a sterile and an active neutrino on BBN has been
studied by various authors \cite{Pe,BaDo,Cl,Sh}. For instance, the
observed helium abundance has led Shi {\it et al.} \cite{Sh} to
exclude large angle sterile neutrino mixing as an MSW
(Mikheyev-Smirnov-Wolfenstein) solution to the solar neutrino
problem.\\

An additional neutrino must interact weakly enough with the $Z^0$ in
order not to violate the constraints from the $Z^0$-decay experiment
at LEP, which requires the total number of active neutrino species to
be $2.993 \pm 0.011$ \cite{LEP}. Moreover, it has to interact weakly
enough not to be counted as a full species during BBN, so that any
additional neutrino species must be sterile. If the sterile neutrinos
were brought into chemical equilibrium before the active neutrinos
freeze out, they would increase the helium yield and bring it in
disagreement with the observed abundance. A sterile neutrino,
$\nu_{\rm s}$, may be produced by helicity flip of active neutrinos.\\

The $\nu_{\rm s}$ will roughly achieve chemical equilibrium if their
production rate $\Gamma_{\rm s}$ is larger than the Hubble constant, $H$,
before the active neutrinos decouple. Hence one can approximately
constrain $\Gamma_{\rm s}$ by requiring that $\Gamma_{\rm s} <H$, i.e.

\begin{equation}
\sigma c\int_{t_{\rm QH}}^{t_{\rm dec}}n(t)\, dt\,\Gamma_{\nu} \leq H ,\label{H}
\end{equation}
where $n(t)$ is the number density of scattering centres; $t_{\rm
dec}$ is the time when neutrinos decouple from the ambient matter,
which is at around 3 MeV for electron neutrinos, and $t_{\rm QH}$ is
the time of the quark-hadron transition at around 100 MeV.\\

The production rate of the active neutrinos is given by

\begin{equation}
\Gamma_{\nu} \simeq\frac{2}{\hbar} G_{\rm F}^2 T^5 ,\label{Gamma}
\end{equation}
where $G_{\rm F}$ is Fermi's constant. Between $t_{\rm dec}$ and $t_{\rm
QH}$ the Universe is almost exclusively composed of photons,
electrons, positrons and neutrinos. Nucleons are rarer than the other
particle species by a factor of about $10^{-10}$. In the temperature
region we are concerned with we can assume that the photons and the
$e^+e^-$ pairs, which interact via the electromagnetic interaction,
are in chemical equilibrium. Since the $e^+e^-$ pairs are far more
abundant than protons, one can neglect the chemical potential of the
electrons (and positrons). Hence, the number density of electrons and
positrons in the early Universe is given by \cite{Peebles}:

\begin{equation}
n_e\simeq\frac{3\zeta (3)}{\pi^2}\left ( \frac{kT}{\hbar c}\right )^3,\label{n}
\end{equation}
where $\zeta$ is the Riemann Zeta-function, $k$ Boltzmann's constant
and $T$ temperature. Substituting Eqs.\,(\ref{sigma}), (\ref{Gamma}) and
(\ref{n}) into Eq.\,(\ref{H}), one finds that for a neutrino
mass of 10 eV and a Hubble constant of 50 km s$^{-1}$ Mpc$^{-1}$:

\begin{equation}
\sigma c\int_{t_{\rm QH}}^{t_{\rm dec}}n_e(t)\, dt\,\Gamma_{\nu} /H
\approx 10^{-45},\label{est}
\end{equation}
i.e. much less than 1, which is the value where BBN would be affected.

\section{Summary}

As shown by Eq.\,(\ref{est}), the effects of spin flips of neutrinos
by torsion of the string theory type are completely insignificant for
the nucleosynthesis predictions. Although the estimate presented in
Eq.\,(\ref{est}) is somewhat crude, it is so small that more
sophisticated calculations are unlikely to yield a significant result.\\

One generalization of the above estimate would be to allow for
nonminimal coupling as suggested by renormalization arguments and some
string theories \cite{Buch}. In the case of nonminimal coupling the
coupling term in the Dirac equation assumes the form
$\frac{C}{4}S_{\mu\nu\sigma}\gamma^{\mu}\gamma^{\nu}\gamma^{\sigma}\psi$,
where $C$ plays the role of an undetermined coupling
constant. However, bounds derived from electron-electron interactions
constrain $C$ to $<1.6\times 10^{14}$ \cite{Ha3}, which is still too
small to raise the expression in Eq.\,(\ref{est}) to a significant
figure.\\

Consequently, theories of gravity that contain torsion of the string
theory type have no noticeable effect on BBN and, conversely, helium
abundance observations cannot serve to place more stringent
constraints on the coupling constant. Thus one can conclude that
quantum gravity theories with a coupling of the kind considered here
are not ruled out as alternative theories of gravity by observations
of the helium abundance in the Universe.

%\acknowledgements

\end{document}